\documentclass[epj]{svjour}
\usepackage{graphics}
\begin{document}
\title{Resonances in chiral unitary approaches}
\author{A. Ramos \inst{1}, V. K. Magas \inst{1},
T. Mizutani \inst{2},
E. Oset \inst{3} and
L. Tol\'os \inst{4}
}                     
%
%
\institute{Departament d'Estructura i Constituents de la
Mat\`eria, Universitat de Barcelona, Diagonal 647, 08028
Barcelona, Spain \and Department of Physics, Virginia Polytechnic
Institute and State University, Blacksburg, VA 24061, USA \and
Departamento de F\'{\i}sica Te\'orica and IFIC, Centro Mixto
Universidad de Valencia-CSIC, Institutos de Investigaci\'on de
Paterna, Apdo. Correos 2085, 46071, Valencia, Spain \and Frankfurt
Institute for Advanced Studies. J.W. Goethe-Universit\"at,
Ruth-Moufang-Str. 1, 60438 Frankfurt am Main, Germany}
\date{Received: date / Revised version: date}
%
\abstract{The extension of chiral theories to the description of
resonances, via the incorporation of unitarity in coupled channels,
has provided us with a new theoretical perspective on the nature
of some of the observed excited hadrons. In this contribution some
of the early achievements in the field of baryonic resonances are
reviewed, the recent evidence of the two-pole nature of the
$\Lambda(1405)$ is discussed and results on charmed baryon
resonances are presented. }

%
\maketitle
\section{Introduction}
\label{intro}

The chiral effective lagrangians describing the interactions of
mesons and baryons respect the basic symmetries of the fundamental
QCD theory. The low energy processes are well described within the
so called chiral perturbation theory ($\chi PT$), which is an
expansion of these lagrangians according to the number of
derivatives of the meson and baryon fields. In recent years, the
introduction of unitarity constraints in coupled channels has led
to extensions of the chiral theories that can be applied at much
higher energies. An interesting consequence of these
non-perturbative methods has been the prediction of many low-lying
baryon and meson resonances that are generated dynamically through
the multiple scattering of their meson or baryon components. The
nature of these states is intrinsically different from the usual
$q\bar q$ or $qqq$ structure of meson and baryons in the sense
that they are not preexistent states that remain in the large
$N_c$ limit where the multiple scattering is suppressed.

In this contribution we give an overview of some of the last
findings in the field of dynamically generated baryon resonances
paying an special attention to the recently confirmed two-pole
nature of the $\Lambda(1405)$. We will also present some
predictions of charm resonances obtained from an extension of the
model to SU(4).

\section{Chiral unitary model}

A compilation of chiral unitary methods can be found in
Ref.~\cite{report}. The search for dynamically generated
resonances proceeds by first constructing the meson-baryon coupled
states from the octet of ground state positive-parity baryons
($B$) and the octet pseudoscalar mesons ($\Phi$) for a given
strangeness channel. The interaction between mesons and baryons is
built from the effective chiral lagrangian which, at lowest order
in momentum, reads
\begin{equation}
L_1^{(B)} = \langle \bar{B} i \gamma^{\mu} \frac{1}{4 f^2} [(\Phi
\partial_{\mu} \Phi - \partial_{\mu} \Phi \Phi) B - B (\Phi
\partial_{\mu} \Phi - \partial_{\mu} \Phi \Phi)] \rangle
\end{equation}
and produces the following driving interaction in $S$-wave
\begin{equation}
{V_{i j}} = - {C_{i j} \frac{1}{4 f^2}(2\sqrt{s} - M_{i}-M_{j})}
\sqrt{\frac{M_{i}+E_i}{2M_{i}}} \sqrt{\frac{M_{j}+E_j}{2M_{j}}}
 \ , \label{eq:v}
\end{equation}
where $i,j$ are indices standing for the different meson-baryon
states in the coupled channel problem, the constants $C_{ij}$ are
SU(3) coefficients encoded in the chiral lagragian and $f$ is the
meson decay constant. The scattering matrix amplitudes between the
various meson-baryon states are then obtained by solving the
coupled channel equation
\begin{equation}
T_{ij} = V_{ij} + V_{il} G_{l} T_{lj} \ , \label{eq:BS}
\end{equation}
where the $V_{il}$ and $T_{lj}$ amplitudes are taken on-shell.
This is a particular case of the N/D unitarization method when the
unphysical cuts are ignored \cite{OO99,OM00}. Under these
conditions the diagonal matrix $G_l$ is simply  built from the
convolution of a meson and a baryon propagator and can be
regularized either by a cut-off ($q_{\rm max}^l$) or,
alternatively, by dimensional regularization depending on a
subtraction constant ($a_l$) coming from a subtracted dispersion
relation.

\section{The $\Lambda(1405)$ and its two-pole nature}

\begin{figure}[h!]
\begin{center}
\resizebox{0.45\textwidth}{!}{%
  \includegraphics{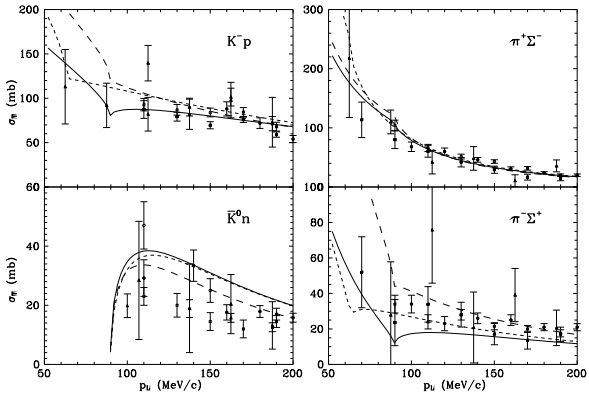}
}
 \caption{ $K^-p$ scattering
cross sections as functions of the $K^-$ momentum in the lab
frame: with the full basis of physical states (solid line),
omitting the $\eta$ channels (long-dashed line) and with the
isospin-basis (short-dashed line). Taken from
Ref.~\protect\cite{OR98}. \label{fig:kncross}}
\end{center}
\end{figure}

The history of the $\Lambda(1405)$ as a dynamical resonance
generated from the interaction of meson baryon components in
coupled channels is long \cite{Jones:1977yk}, but it has
experienced a boost within the context of unitary extensions of
chiral perturbation theory
\cite{Kai95,OR98,Oller:2000fj,Lutz:2001yb,Garcia-Recio:2002td,ORB02},
where the lowest order chiral Lagrangian and unitarity in coupled
channels generates the $\Lambda(1405)$  and leads to good
agreement with the $K^- p$ reactions. The models have been fine
tuned recently by including the additional terms of the
next-to-leading order lagrangian
\cite{Oller:2005ig,Borasoy:2005ie,Borasoy:2006sr,Oller:2006jw}.
The pioneer work of Ref.~\cite{Kai95} included the channels
closest to the $\Lambda(1405)$, namely $\pi\Sigma$ and $\bar K N$
in $I=0$ and $\pi\Lambda$, $\pi\Sigma$ and $\bar K N$ in $I=1$,
obtaining a good reproduction of the scattering observables only
if the next-to-leading order lagrangian was also considered. In
the work of Ref.~\cite{OR98} all the channels that can be built
form the lowest lying pseudoscalar mesons and the octet of ground
state baryons were employed, namely $\bar{K}^0 n$, $\pi^0
\Lambda$, $\pi^0 \Sigma^0$, $\pi^+ \Sigma^-$, $\pi^- \Sigma^+$,
$\eta \Lambda$, $\eta \Sigma^0$, $K^+ \Xi^-$, $K^0 \Xi^0$ in the
case of $K^- p$ scattering. It was noted that, even if the
additional channels included are well above the $K^- p$ threshold,
there are important interferences between the real parts of the
amplitudes and these channels, especially the $\eta \Lambda$ and
$\eta \Sigma^0$ ones, were extremely important to reproduce the
threshold branching ratios taking simply the chiral lagrangian at
lowest order. This is clearly illustrated in
Fig.~\ref{fig:kncross} were the elastic and inelastic $K^- p$
cross section are compared with the results from the model of
Ref.~\cite{OR98}.

\vspace*{0.3cm}

The surprise, however, came with the realization that there are
two poles in the neighborhood of the $\Lambda(1405)$ both
contributing to the final experimental invariant mass distribution
\cite{Oller:2000fj,Garcia-Recio:2002td,Jido:2002yz,Jido:2003cb,Garcia-Recio:2003ks,Hyodo:2002pk,Nam:2003ch}.
The properties of these two states are quite different, one has a
mass around $1390$ MeV, a large width of about $130$ MeV and
couples mostly to $\pi \Sigma$, while the second one has a mass
around $1425$ MeV, a narrow width of about $30$ MeV and couples
mostly to $\bar{K} N$. The  two states are populated with
different weights in different reactions and, hence, their
superposition can lead to different distribution shapes. Since the
$\Lambda(1405)$ resonance is always seen from the invariant mass
of its only strong $\pi \Sigma$ decay channel, hopes to see the
second pole are tied to having a reaction where the
$\Lambda(1405)$ is formed from the $\bar{K} N$ channel.  In this
sense a calculation of the $K^- p \to \gamma \Lambda(1405)$
reaction \cite{Nacher:1999ni}, prior to the knowledge of the
existence of the two $\Lambda(1405)$ poles, showed a narrow
structure at about 1420 MeV.  With the present perspective this is
clearly interpreted as the reaction proceeding through the
emission of the photon followed by the generation of the resonance
from $K^- p$, thus receiving a large contribution from the second
narrower state at higher energy.
The recently measured reaction $K^- p \to \pi^0 \pi^0 \Sigma^0$
\cite{Prakhov} allows us to test already the two-pole nature of
the $\Lambda(1405)$. This process shows a strong similarity with
the reaction $K^- p \to \gamma \Lambda(1405)$, where the photon is
replaced by a $\pi^0$.

A model for the reaction $K^- p \to \pi^0 \pi^0 \Sigma^0 $ in the
energy region of $p_{K^-}=514$ to $750$ MeV/c, as in the
experiment \cite{Prakhov}, has been studied in
Ref.~\cite{Magas:2005vu}. The mechanisms considered are those in
which a $\pi^0$ loses the necessary energy to allow the remaining
$\pi^0\Sigma^0$ pair to be on top of the $\Lambda(1405)$
resonance.  It was found that the dominant contribution was the
nucleon pole term, in which the $\pi^0$ is emitted from the proton
and, consequently, the resonance is initiated by a $K^- p$ state.
The $\Lambda(1405)$ thus obtained comes mainly from the $K^- p \to
\pi^0 \Sigma^0$ amplitude which, as mentioned above, gives  the
largest possible weight  to the second (narrower) state.

\begin{figure}[htb]
\begin{center}
\hspace*{-0.5cm} \resizebox{0.55\textwidth}{!}{%
\includegraphics{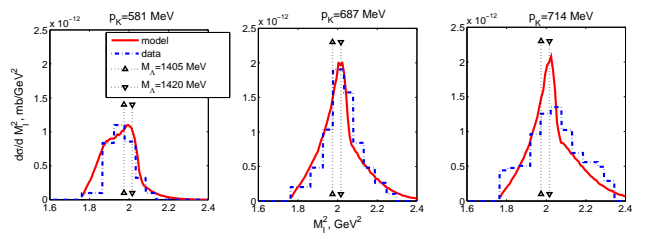} } \caption{The $\pi^0
\Sigma^0$ invariant mass distribution for three different initial
kaon momenta. Taken from Ref.~\protect\cite{Magas:2005vu}.}
\end{center}
\label{fig:mass}
\end{figure}
%

%



\begin{figure}[tb]
\begin{center}
\resizebox{0.35\textwidth}{!}{%
  \includegraphics{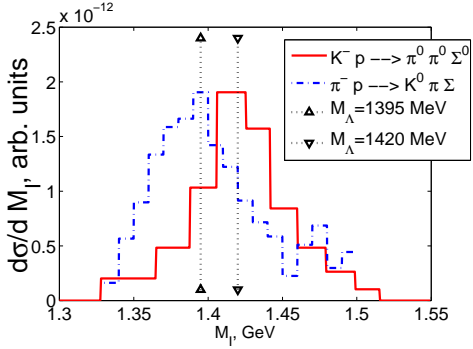}
}
 \caption{Two experimental shapes of  $\Lambda(1405)$ resonance.
 See text for more details. Taken from
 Ref.~\protect\cite{Magas:2005vu}.}
 \label{two_exp}
\end{center}
\end{figure}


In Fig.~2 the $\pi^0\Sigma^0$ invariant mass distribution obtained
in Ref.~\cite{Magas:2005vu} for three different energies of the
incoming $K^-$ are compared to the experimental data.
Symmetrization of the amplitudes produces a sizable amount of
background. At a kaon laboratory momentum of $p_K=581$ MeV/c this
background  distorts the $\Lambda(1405)$ shape producing cross
section in the lower part of $M_I$, while at $p_K=714$ MeV/c the
strength of this background is shifted toward the higher $M_I$
region. An ideal situation is found for momenta around $687$
MeV/c, where the background sits below the $\Lambda(1405)$ peak
distorting its shape minimally. The peak of the resonance shows up
at $M_I^2=2.02$ GeV$^2$ which corresponds to $M_I=1420$ MeV,
larger than the nominal $\Lambda(1405)$, and in agreement with the
predictions of Ref.~\cite{Jido:2003cb} for the location of the
peak when the process is dominated by the $t_{{\bar K}N \to
\pi\Sigma}$ amplitude.  The apparent width from experiment is
about $40-45$ MeV, but a precise determination would require to
remove the background mostly coming from the ``wrong'' $\pi^0
\Sigma^0$ pairs due to the indistinguishability of the two pions.
The theoretical analysis of Ref.~\cite{Magas:2005vu} permits
extracting the pure resonant part by not symmetrizing the
amplitude, finding a width of $\Gamma=38$ MeV, which is smaller
than the nominal $\Lambda(1405)$ width of $50\pm 2$ MeV
\cite{PDG}, obtained from the average of several experiments, and
much narrower than the apparent width of about $60$ MeV that one
sees in the $\pi^- p \to K^0 \pi \Sigma$ experiment \cite{Thomas},
which also produces a distribution peaked at $1395$ MeV. In order
to illustrate the difference between the $\Lambda(1405)$ resonance
seen in this latter reaction and in the present one, the two
experimental distributions are compared in Fig.~\ref{two_exp}. We
recall that the shape of the $\Lambda(1405)$ in the $\pi^- p \to
K^0 \pi \Sigma$ reaction was shown in Ref.~\cite{hyodo} to be
largely built from the
 $\pi \Sigma \to \pi \Sigma$ amplitude, which is dominated by
the wider, lower energy state.

 We have thus seen that the $K^- p \to
 \pi^0 \pi^0 \Sigma^0$ reaction is particularly suited to study the features
 of the second pole of the $\Lambda(1405)$ resonance, since it is largely
 dominated by a mechanism in which a $\pi^0$ is emitted prior to the $K^- p \to
 \pi^0 \Sigma^0$ amplitude, which is the one giving the largest weight to the
 second narrower state at higher energy.
 The model of Ref.~\cite{Magas:2005vu} has proved to be accurate in reproducing both the
 invariant mass distributions and  integrated cross sections seen in
 a recent experiment \cite{Prakhov}.  The study of the $K^- p \to
 \pi^0 \pi^0 \Sigma^0$
 reaction, complemental to the one of Ref.  \cite {hyodo} for the $\pi^- p \to
 K^0 \pi \Sigma$ reaction, has shown that the quite different shapes of the
 $\Lambda(1405)$ resonance seen in these experiments can be interpreted in favour
 of the existence of two poles with the corresponding states having the
 characteristics predicted by the chiral theoretical calculations.
 Besides demonstrating once more the great predictive power of the chiral
 unitary theories, this combined study of the two reactions gives the first
 clear evidence of the two-pole nature of the $\Lambda(1405)$.

Various other reactions testing the two-pole nature of the
$\Lambda(1405)$ have been presented at this workshop
\cite{oset-here}.

\section{Extensions within the flavour SU(3) sector}

  The strangeness $S= 0$ channel was also investigated using the Lippmann-
Schwinger equation and coupled channels in
Ref.~\cite{Kaiser:1995cy,Kaiser:1996js}. The $N^*(1535)$ resonance
was found to be generated dynamically within this approach.
Subsequently, work was done in this sector
\cite{Nacher:1999vg,Nieves:2001wt}, and the $N^*(1535)$ resonance,
as well as the low-energy scattering observables, were well
reproduced, with the exception of the isospin 3/2 channel.
 Ref.~\cite{Inoue:2001ip} continued and further improved work along these lines
  by introducing the $\pi N \to \pi NN$ channels, which proved essential in
 reproducing the isospin 3/2 part of the $\pi N$ amplitude.

The $S=-2$ sector was investigated within a chiral unitary theory
in the work of Ref.~\cite{Ramos:2002xh}. By taking  the parameters
of the model to be of natural size, a pole in the amplitude was
found at $1605 + {\rm i}65$ MeV, which coupled strongly to
$\pi\Xi$ and $K\Lambda$ states and very mildly to the $K\Sigma$
and $\eta\Xi$ channels. While the width of this resonance might
appear as grossly overestimating that of the experimentally
observed $\Xi(1620)$ and $\Xi(1690)$ resonances, of around 50 MeV,
the calculated
 $\pi \Xi$ invariant mass distribution shows a smaller apparent width  compared to
the one obtained at the pole position, due to the fact that the
resonance appears just below the threshold of the  $\bar{K}
\Lambda$ channel to which it couples very strongly (Flatt\'e
effect). The generated resonance was identified with the
$\Xi(1620)$, because the $\Xi(1680)$ has branching ratios for
$\bar{K}\Sigma$ to $\bar{K}\Lambda$ of around 3 and for $\pi \Xi$
to $\bar{K}\Sigma$ of less than 0.09, values that are incompatible
with the properties obtained in the model of
Ref.~\cite{Ramos:2002xh}.  Therefore, it was possible to assign
the unknown spin and parity of this resonance to $J^P=1/2^-$,
which are the values of the states produced with the meson-baryon
$S$-wave lagrangians. These findings were confirmed in
Ref.~\cite{Garcia-Recio:2003ks}, were a second $S=-2$ pole that
could be identified with the $\Xi(1680)$ resonance was also found.

Recently, a good amount of work has been devoted to the study of
$J^P=3/2^-$ states that can be dynamically generated through the
interaction of pseudoscalar $0^-$ mesons with decuplet $3/2^+$
baryons
\cite{Kolomeitsev:2003kt,Sarkar:2005ap,Roca:2006sz,Hyodo:2006uw,Doring:2006ub,Doring:2005bx,Doring:2006pt},
and including also the vector meson degrees of freedom within spin-flavor SU(6)
symmetry \cite{Garcia-Recio:2005hy}.

The $S=-1$ and $I=0$ sector forms a two-coupled $\pi \Sigma^*$ and
$K \Xi^*$ problem that generates a real pole close to the mass of
the observed $\Lambda(1520)$. Obtaining a width is only possible
if one also includes lower-lying channels formed by a pseudoscalar
meson and a baryon of the $1/2^+$ octet, such as $\bar{K}N$ and $
\pi \Sigma$, interacting in $D$-wave with couplings fitted to
experimental amplitudes \cite{Sarkar:2005ap,Roca:2006sz}. The
cross sections of the model for the $K^-p\to\pi^0\pi^0\Lambda$ and
$K^-p\to\pi^+\pi^-\Lambda$ reactions compare very satisfactory
with data \cite{Prakhov:2004ri,Mast:1973gb}. This has to be
considered as a non-trivial accomplishment of the theory  since
the $\bar{K} N \to \pi \Sigma^*$ amplitude has not been included
in the fit. Other studies related to the $\Lambda(1520)$ comprise
the coupling of the resonance to $\bar K^* N$ states
\cite{Hyodo:2006uw} and its radiative decays \cite{Doring:2006ub}.

In the $S=0$ $I=3/2$ sector, the $\Delta(1700)$ is another of the
resonances that are obtained from the interaction of the octet of
mesons with the baryon decuplet
\cite{Doring:2005bx,Doring:2006pt}. The coupling of this resonance
to its $\pi\Delta$, $K\Sigma^*$ and $\eta\Delta$ components are
found to be very different than those obtained if the
$\Delta(1700)$ was assumed to belong to a SU(3) decuplet, as
suggested by the PDG \cite{PDG}. Assuming one or the other picture
would give rise to very different predictions for cross sections
in reactions involving the excitation of the $\Delta(1700)$.
Therefore, the agreement of the theory with data for pion and
photon induced reactions \cite{Doring:2006pt} must again be
considered as a success of the chiral unitary models.

\section{Dynamical resonances with charm}

The physics in the charm $C=1$ sector bears a strong resemblance
with the phenomenology seen in $\bar K N$ dynamics, once the $s$
quark is replaced by a $c$ quark. This is reinforced by an
apparent correspondence between the two $I=0$ $S$-wave
$\Lambda(1405)$ and $\Lambda_c(2593)$ resonances. The mass of the
former lies in between the thresholds of the $\pi\Sigma$ and $\bar
K N$ channels, to which it couples strongly. The $\Lambda_c(2593)$
lies below the $DN$ threshold and just slightly above the
$\pi\Sigma_c$ one. It is therefore logical to explore whether the
$\Lambda_c(2593)$ can also be dynamically generated from the
interaction of $DN$ pairs and its related channels. The study of
the $DN$ interaction is also interesting by itself, in connection
to the behavior of $D$ mesons in a nuclear medium for the
implications it may have on the phenomenon of $J/\Psi$
suppression, which is connected with the formation of a
quark-gluon plasma in relativistic heavy-ion collisions, and on
the production of open-charm mesons that will be explored in the
CBM experiment at the future FAIR facility \cite{CBM}.

The in-medium modifications of $D$ mesons from a coupled channel
perspective was first explored in Ref.~\cite{Tolos:2004yg}, where
free space amplitudes were constructed from a set of separable
coupled-channel interactions obtained with chirally motivated
lagrangians upon replacing the $s$ quark by the $c$ quark,
 to reproduce the $I=0$ $\Lambda_c(2593)$ as a
$DN$ $S$-wave {\it hadronic molecular} state of binding energy
$\approx 200$ MeV with a width of $\approx 3$ MeV, sitting very
close to the $\pi \Sigma_c$ threshold. This work represented the
first indication that the  $\Lambda_c(2593)$ could have a
dynamical origin but, by ignoring the strangeness degree of
freedom, the role of the $D_s \Lambda$, $D_s \Sigma$, $K\Xi_c$ and
$K\Xi_c^\prime$ states was excluded from this model. In addition,
both $\pi$ and $K$ Goldstone mesons should be considered on an
equal footing within chiral symmetry, while, on the other hand,
the $D$ mesons are much heavier and obey heavy-quark symmetry. A
blind $s\to c$ replacement breaks both of those symmetries.

A different approach, which respected the proper symmetries, was
attempted in Ref.~\cite{Lutz:2003jw}. There, charmed baryon
resonances were generated dynamically from the scattering of
Goldstone bosons off ground-state $J^P=\frac{1}{2}^+$ charmed
baryons. The $C=1$, $S=I=0$ resonance found at 2650 MeV was
identified with the $\Lambda_c(2593)$ in spite of the fact that
the width, due to the strong coupling to $\pi\Sigma_c$ states,
came out to be more than twenty times larger than the experimental
width of about 4 MeV. The problem of this model is that, apart
from ignoring the $D_s Y$ states, it does not account either for
the coupling to the $DN$ channel, which contributes strongly to
generating the narrow $\Lambda_c(2593)$ according to
Ref.~\cite{Tolos:2004yg}.

  A substantial improvement came in a recent work
\cite{Hofmann:2005sw} in which the alleged  shortcomings  have
been overcome by exploiting the universal vector meson coupling
hypothesis to break the $SU(4)$ symmetry in a convenient and
well-defined  manner. More precisely, this is done by a
$t$-channel exchange of vector mesons between pseudoscalar mesons
and baryons in such a way  to respect  chiral symmetry for the
light meson sector and the heavy quark symmetry for charmed
mesons,  as well as to maintain the interaction to be of the
Tomozawa-Weinberg (T-W) vector type. The model generates a narrow
$C=1$, $I=0$ resonance that it is identified with the
$\Lambda_c(2593)$, together with an almost degenerate $S$-wave
resonance at $2620$ MeV in the $C=1$, $I=1$ channel, which is not seen
experimentally. An application of this model to a
preliminary study of $D$ and $D_s$ mesons in nuclear matter may be
found  in Ref. \cite{Lutz:2005vx}.

 The work of Ref.~\cite{Mizutani:2006vq} implemented
some modifications over the model developed in
Ref.~\cite{Hofmann:2005sw}. In the first place, the $t$-dependence
of the interaction kernel was eliminated and only the leading
order zero-range terms were considered. This was motivated from
the observation \cite{Mizutani:2006vq} that the $t\to 0$ limit
implemented in the energy denominator of the kernel compensated to
a good extent the term $k_\mu k_\nu/m_X^2$ in the numerator, which
was retained in Ref.~\cite{Hofmann:2005sw}. The resulting
interaction was of the form shown in Eq.~\ref{eq:v}, but with an
additional factor 1/4 for charm-exchange transitions, accounting
approximately for the ratio $(m_V/m_V^c)^2$ between the squared masses 
of uncharmed and charmed vector mesons, such as the $\rho$ and $D^*$,
respectively. Another novel feature of the model of
Ref.~\cite{Mizutani:2006vq} is that the T-W interaction is
supplemented with a scalar-isoscalar attraction characterized by a
$\Sigma_{DN}$ term, which, due to the large difference between the
$c$ and $s$ quark masses, might be more relevant in the charm
sector than in the strange one. Mean-field approaches have shown
that this term has a tremendous influence on the in-medium $D$ and
$\bar D$ properties \cite{TSU99,SIB99,MIS04}. Finally, in view of
the application of the model to determining the $D$ meson
properties in nuclear matter, a cut-off regularization scheme was
implemented in Ref.~\cite{Mizutani:2006vq}. The value of the
cut-off was adjusted to reproduce the position and width of the
$\Lambda_c(2593)$ resonance, allowing also for a fine-tuning of
the strength of the Goldstone boson decay constant $f$, as in
Ref.~\cite{OR98}. Two models were explored, the difference being
the inclusion or not of the $\Sigma_{DN}$ term. The corresponding
parameters are, model A: $f=1.15f_{\pi},\
\Sigma=\Sigma_{DN}/f_D^2=0.05$ MeV$^{-1}$, $\Lambda=727$ MeV, and
model B: $f=1.15f_{\pi},\ \Sigma=\Sigma_{DN}/f_D^2=0$ MeV$^{-1}$,
$\Lambda=787$ \ MeV, where $\Lambda$ is the ultra-violet cut-off
value for the integration in the loop $G_l$.

\begin{figure}[htb]
\begin{center}
\resizebox{0.47\textwidth}{!}{%
  \includegraphics{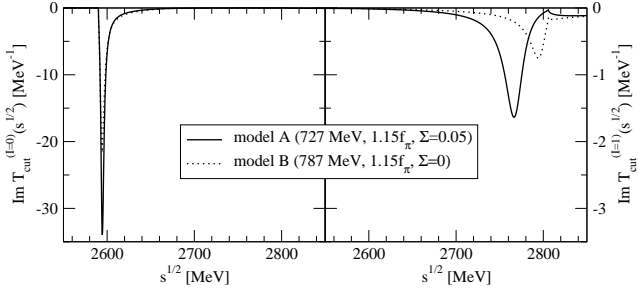}
}
\caption{Imaginary part of the $DN$ amplitude as
a function of $\sqrt{s}$ for $I=0$ (left panel) and $I=1$ (right
panel), obtained by using the cut-off scheme with model A (solid
lines) and model B (dotted lines). Taken from
Ref.~\protect\cite{Mizutani:2006vq}.} \label{T_cut_2}
\end{center}
\end{figure}

As seen on the left panel of Fig.~\ref{T_cut_2}, the width of the
$I=0$ resonance is found to be $\sim 4$ MeV for model A and $\sim
5$ MeV for model B, respectively.  In the $I=1$ sector, shown in
the right panel of the figure, both models generate a resonance
close to 2800 MeV, with a width of around 30 MeV. We note that the
Belle Collaboration has recently measured in this energy range an
isotriplet of excited charmed baryons decaying into
$\Lambda_c^+\pi^-$, $\Lambda_c^+\pi^0$ and $\Lambda_c^+\pi^+$
\cite{belle05}.  It is interpreted as a new charmed baryon, the
$\Sigma_c(2800)$, having a width of around 60 MeV, measured with
more than 50 \% error.  This baryon has been tentatively
identified with a $D$-wave resonance, to conform to quark model
predictions \cite{Copley79}, although the predicted width $\Gamma
\sim 15$ MeV \cite{Pirjol97} is substantially smaller than the
observed one. Actually, the fits performed in \cite{belle05} were
not too sensitive to varying the signal parameterization using
$S$-wave or $P$-wave Breit-Wigner functions, hence this state
could still qualify as $S$-wave type and, therefore, according
to results of the charm-extended chiral unitary model of
Ref.~\cite{Mizutani:2006vq}, it could be interpreted as being a
dynamically generated resonance.

Models have already been extended to the study of $D$-wave charm
resonances through the interaction of Goldstone mesons with
decuplet baryons \cite{Hofmann:2006qx}, or considering the
additional role of the interaction of vector mesons with octet and
decuplet baryons \cite{granada}.

We finalize this section by showing the properties of $D$ mesons
in a hot nuclear medium, as obtained in Ref.~\cite{Tolos:2007vh}
using the above described dynamical model for the $DN$ interaction
\cite{Mizutani:2006vq}. The in-medium solution at finite
temperature incorporates Pauli blocking effects, mean-field
binding on all the baryons involved, and $\pi$ and open-charm
meson self-energies in a self-consistent manner.

\begin{figure}[htb]
\begin{center}
\resizebox{0.47\textwidth}{!}{%
  \includegraphics{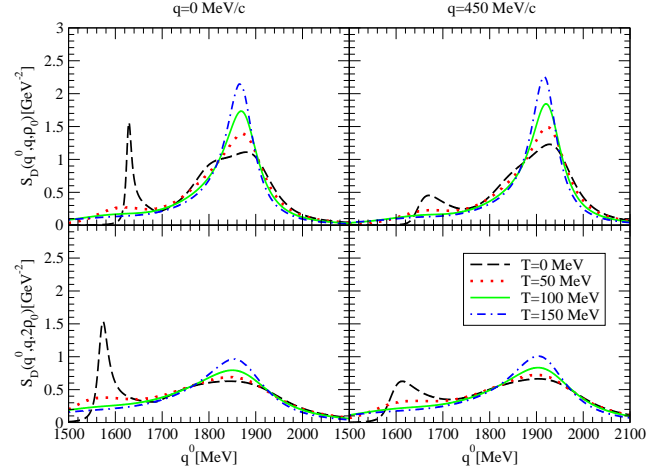}
}
\caption{The evolution of the $D$ meson spectral
function with temperature for $\rho=\rho_0$ (upper panels) and
$\rho=2\rho_0$ (lower panels). Taken from
Ref.~\protect\cite{Tolos:2007vh}.} \label{fig:spec_evol}
\end{center}
\end{figure}

 The obtained evolution of the $D$-meson spectral function
with temperature is seen in Fig.~\ref{fig:spec_evol} for two
densities, $\rho_0$ and $2\rho_0$, and two momenta, $q=0$ MeV/c
and $q=450$ MeV/c, for model A. Two peaks appear in the spectral
density at $T=0$. The lower one corresponds to the $\tilde
\Lambda_c N^{-1}$ excitations, whereas the  higher one is the
quasi-particle peak mixed with  the $\tilde \Sigma_c N^{-1}$ mode.
The position of the quasi-particle peak lies not too far away from
the free mass, a feature which is in strong contrast with the
large attractive shifts found in mean-field models
\cite{TSU99,SIB99,MIS04}. The $\tilde \Lambda_c N^{-1}$ and
$\tilde \Sigma_c N^{-1}$  structures dissolve with increasing
temperature, while the quasi-particle peak becomes narrower and
moves closer to its free value position. This is due to the fact
that the self-energy receives contributions from $DN$ pairs at
higher momentum where the interaction is weaker. Similar features
were observed in an earlier study \cite{Tolos:2005ft}, which used
the three-flavour dynamical model developed in
Ref.~\cite{Tolos:2004yg}.

The widening of the quasi-particle peak for larger nuclear density
may be understood as due to enhancement of collision and
absorption processes. The $\tilde\Lambda_c N^{-1}$ mode moves down
in energy with increasing density due to the lowering in the
position of the $\tilde\Lambda_c$ resonance induced by the more
attractive $\Sigma_{DN}$ term in model A.

The properties of the $\bar D$ meson in hot nuclear matter
were also investigated in
the work of Ref.~\cite{Tolos:2007vh}, where it was found that
its mass develops a mild repulsive shift and that, despite the
absence of resonant structures in the $\bar D N$ interaction, the
low-density approximation to the repulsive $\bar D$ self-energy
was not reliable even at subsaturation densities.

Taking into account these results, one can already state that, while $D^-$
-mesic atoms will always be bound by the Coulomb interaction, no strongly bound
nuclear states or even bound $\bar{D}^0$ nuclear systems are expected due to
the repulsive in-medium mass shift obtained. In the charm $C=1$ sector, an
experimental observation of bound $D$ nuclear states is ruled out by the
moderate attraction and large width found for the $D$ meson. With respect to
the implications on $J/\Psi$ suppression in hadronic reactions, we note that
the process $J/\Psi \to D \bar D$ cannot proceed spontaneously in free space
due to a mass difference of $\approx 650$ MeV. According to the self-consistent
models in nuclear matter, the in-medium $\bar D$ mass is seen to increase by
about $10-20$ MeV whereas the tail of the quasi-particle peak of the $D$
spectral function extends to lower energies due to the thermally spread $\tilde
Y_c N^{-1}$ configurations. Nevertheless, it is very unlikely that this lower
tail extends as far down by more than 600 MeV with sufficient strength to give
rise to a direct disappearance of $J/\Psi$'s. Alternatively, $J/\Psi$
suppression in a hadronic environment could also proceed 
by cutting its supply from the
excited charmonia: $\chi_{c\ell}(1P)$ or $\Psi'$. These states will be strongly
absorbed in the medium by multi-nucleon processes, which will be enhanced by
the low-energy tail of the spread out $D$-meson spectral function.

\section*{Acknowledgments}

This work is partly supported by contracts 
RII3-CT-2004-506078 \hfil (HadronPhysics, EU),\hfil
MRTN-CT-2006-035482 \\ 
\noindent (FLAVIAnet, EU), \hfil
FIS2005-03142 \hfil and \hfil FIS2006-03438\\
\noindent (MEC, Spain), 2005SGR-00343
(Generalitat de Ca\-ta\-lu\-nya) and \hfil ANBest-P \hfil and\hfil BNBest-BMBF
\hfil 98/NKBF98 \\
\noindent (Germany).
%
%

\end{document}